\documentclass[review]{elsarticle}

\usepackage{lineno,hyperref}
\usepackage[normalem]{ulem}
\usepackage{graphics}
\usepackage{amsmath,bm}
\usepackage{amssymb}
\usepackage{graphicx}
\usepackage{bm}
\usepackage{amsfonts}
\usepackage{color}

\usepackage{epstopdf}
\usepackage{booktabs}
\usepackage{array}
\usepackage{graphics}
\usepackage[section]{placeins}

\renewcommand{\vec}[1]{\ensuremath{\mathbf{#1}}}

\newcommand{\pdev}[1]{\ensuremath{\partial_{#1}}}

\newcommand*{\vcenteredhbox}[1]{\begingroup       
\setbox0=\hbox{#1}\parbox{\wd0}{\box0}\endgroup}

\modulolinenumbers[5]
\journal{Journal of Computational Science}

\begin{document}

\begin{frontmatter}

\title{Sub-Kolmogorov droplet dynamics in isotropic turbulence using a multiscale Lattice Boltzmann scheme}


\author[address1,address2]{Felix Milan\corref{mycorrespondingauthor}}
\cortext[mycorrespondingauthor]{Corresponding author}
\ead{felix.milan@roma2.infn.it}

\author[address1]{Luca Biferale}
\ead{biferale@roma2.infn.it}

\author[address1]{Mauro Sbragaglia}
\ead{sbragaglia@roma2.infn.it}

\author[address2,address3,address4]{Federico Toschi}
\ead{f.toschi@tue.nl}

\address[address1]{Department of Physics and INFN, University of Rome ``Tor Vergata'', Via della Ricerca Scientifica 1, 00133 Rome, Italy}
\address[address2]{Department of Applied Physics, Eindhoven University of Technology, Eindhoven 5600 MB, The Netherlands}
\address[address3]{Department of Mathematics and Computer Science, Eindhoven University of Technology, Eindhoven 5600 MB, The Netherlands}
\address[address4]{CNR-IAC, Via dei Taurini 19, 00185 Rome, Italy}

\begin{abstract} The deformation and dynamics of a single droplet in isotropic turbulence is studied using a Lattice Boltzmann diffuse interface model involving exact boundary flow conditions~\cite{Milan2018} to allow for the creation of an external turbulent flow. We focus on a small, sub-Kolmogorov droplet, whose scale is much smaller than the Kolmogorov length scale of the turbulent flow. The external flow field is obtained via pseudo-spectral simulation data describing the trajectory of a passive tracer in isotropic turbulence. In this way we combine the microscopic scale of the droplet and the macroscopic scale of the turbulent flow. The results obtained from this fully resolved model are compared to previous studies on sub-Kolmogorov droplet dynamics in isotropic turbulent flows~\cite{Biferale2014}, where an analytical model~\cite{MaffettoneMinale98}, which assumes the droplet shape to be an ellipsoid at all times, is used to describe the droplet deformation. Our findings confirm, that the hybrid pseudo-spectral Lattice Boltzmann algorithm is able to study droplet deformation and breakup in a regime well beyond the ellipsoidal approximation.
\end{abstract}


\end{frontmatter}

\section{Introduction}

The dynamics and breakup behaviour of immiscible droplets in laminar flows have been studied extensively in the literature, from applied studies involving concrete setups~\cite{Flumerfelt72,Singh2010,Shewan2013,Rodriguez2015} to theoretical studies addressing their physical complexity~\cite{Fortelny2019,Taylor1932,Greco02,GuidoRev,Janssen08}. A lot of attention has been dedicated to droplet deformation and breakup in stationary laminar flows~\cite{Fortelny2019,Acrivos1964,Lyngaae1990,Vananroye08,Vananroye11b,Guido11,Sibilloetal06,Renardy2001,Renardy08,GuidoRev,Greco02,Janssen10,Milan2020}, but less is known about the dynamics of droplets in homogeneous isotropic turbulent flows, which pose the challenge of solving a multi-physics problem, since the flow properties of the turbulent scales have to be accurately transferred to the scale of the droplet. We can differentiate between two interesting cases for the deformation of droplets in homogeneous and isotropic turbulent flows~\cite{Elghobashi2019}: On the one hand, there are studies on droplet dynamics on scales larger than the Kolmogorov scale~\cite{Albernaz2017,Njobuenwu2015,Komrakova2015,Komrakova2019,Mukherjee2019}. On the other hand, we investigate the dynamics and breakup statistics of sub-Kolmogorov droplets, i.e. droplets whose size is much smaller than the Kolmogorov scale~\cite{Cristini03,Khismatullin03,Komrakova13,Biferale2014,Spandan2016,Ray2018}. The latter case is of particular interest to the authors, since the viscous stresses dominate over the inertial stresses, if the droplet size is smaller than the Kolmogorov scale of the outer flow~\cite{Cristini03}. In this regime the droplet is deformed similarly to the laminar case and the turbulence is only noticeable via the time-dependency of the solvent flow. Both~\cite{Biferale2014} and~\cite{Ray2018} make use of the Maffettone-Minale model (MM-model)~\cite{MaffettoneMinale98} to model the droplet in homogeneous isotropic turbulence, which only accounts for ellipsoidal deformations and thus can model breakup only via a cut off criteria. Even though there exist several computational studies on sub-Kolmogorov droplet dynamics and breakup in turbulent flows~\cite{Cristini03,Khismatullin03,Komrakova13,Biferale2014,Spandan2016,Ray2018} there exists no experimental evidence of sub-Kolmogorov droplet break up~\cite{Elghobashi2019}, as such experiments are difficult to set up. However, there are experimental studies on breakup in flows through dilute random beds of fixed fibers~\cite{Patel03} and breakup of a small single droplet in a chaotic advective flow~\cite{Tjahjadi91}. These results~\cite{Patel03,Tjahjadi91} show that small droplets, whose size is comparable to the size of a sub-Kolmogorov droplet in a realistic turbulent flow, can break up in highly unsteady and time-dependent laminar flows, indicating that breakup of sub-Kolmogorov droplets is indeed possible and not merely a computational artefact. Moreover, our studies employ a novel multiscale algorithm, as we not only use the MM-model to measure the deformation of the droplet but compare it with fully resolved LBM (Lattice Boltzmann Method) simulations using a diffuse interface method, the Shan-Chen multicomponent model (SCM)~\cite{Shan93,Shan94}. LBM in conjunction with SCM has been extensively used in the field of microfluidics and droplet breakup~\cite{Komrakova13,Liu,Farokhirad,Onishi2,Xi99,Yoshino08,GuptaSbragaglia2014,GuptaSbragagliaScagliarini2015,Milan2020,Chiappini2018,Chiappini2019} and notably also for multicomponent systems involving thermal fluctuations \cite{Varnik11,Xue2018}. But how can we couple LBM droplet dynamics to fully developed turbulence? One way in doing so is to use a LBM hybrid boundary scheme, which is able to simulate an exact turbulent flow at the domain boundary. Exact flow boundary conditions for LBM are a very useful tool for multiscale physics simulations, as it is enabling scale separation in its very own nature. Exact flow boundaries for LBM were first developed by~\cite{Zou97} for 2D lattices and then extended to 3D flows by~\cite{Mattila09,Kutay06,Mattila09b} for perpendicular inlet and outlet flows. In~\cite{Hecht10} it is shown that exact boundary flow conditions can be implemented in 3D LBM simulations for the D3Q19 lattice. Even though the scheme presented in~\cite{Hecht10} seems rather suitable in modelling a sub-Kolmogorov droplet in homogeneous isotropic turbulence, we use a newly developed ghost boundary flow method. The main reason for this is that the method in~\cite{Hecht10} cannot determine both the density $\rho_b(\vec{x},t)$ and the velocity field $\vec{v}(\vec{x},t)$ at the lattice domain boundary. Since we require a constant density ratio of $\rho_{\tiny \mbox{d}} / \rho_{\tiny \mbox{s}}$, where the subscripts $d$ and $s$ denote the droplet and solvent phase respectively, for our hybrid simulations, we use the ghost node boundary scheme, which does not set constraints to either the density nor the velocity field at the boundary. Our numerical method can be seen as a novel hybrid approach to the sub-Kolmogorov droplet dynamics: The simulation domain can be of the same order as the droplet's size, see section~\ref{sec:validation}, but should be ideally much larger than the droplet radius $R$, see section~\ref{sec:results}. The surrounding turbulent flow is used as an input via an open boundary condition method~\cite{Milan2018}, similar to the setups in~\cite{Mattila09,Hecht10}. This enables us to study the deformation and breakup behaviour of a sub-Kolmogorov droplet in fully developed isotropic turbulence, which we can compare to the results obtained via the widely used MM-model. We find that the fully resolved PS LBM hybrid simulation scheme is crucial to model sub-Kolmogorov droplet breakup in homogeneous isotropic turbulence, since it allows for non-ellipsoidal deformations close to breakup in contrast to the MM-model. 

\section{The MM-model and the Lattice Boltzmann algorithm}
\label{sec:method}
The phenomenological MM-model describes the coupling of droplet dynamics to a solvent flow. The droplet is always assumed to be ellipsoidal, so that we can describe it via a second rank tensor $M_{ij}$, which is also referred to as the morphology tensor. Droplet deformation is characterised via the components of $M_{ij}$ (e.g. for an undeformed droplet $M_{ij} = \delta_{ij}$). The time evolution of $M_{ij}$ due to an external flow field is given by the MM equation~\cite{MaffettoneMinale98}:

\begin{align}
\label{eq:mm_general}
\frac{d M_{ij}}{dt} & = \mbox{Ca} \left [ f_2 (S_{ik} M_{kj} + M_{ik} S_{kj}) + \Omega_{ik} M_{kj} - M_{ik} \Omega_{kj} \right ] \notag \\
& - f_1 \left ( M_{ij} - 3 \frac{III_M}{II_M} \delta_{ij} \right )
\end{align}

where $S_{ij}$ and $\Omega_{ij}$ are the symmetric and anti-symmetric parts of the velocity gradient $G_{ij} \equiv \pdev{j} v_i$ respectively:

\begin{align}
\label{eq:shear_tensor}
S_{ij} & = \frac{1}{2} (\pdev{j} v_i + \pdev{i} v_j) \notag \\
\Omega_{ij} & = \frac{1}{2} (\pdev{j} v_i - \pdev{i} v_j)
\end{align}

$III_M = \det (M_{ij})$ and $II_M \equiv \frac{1}{2} (M_{kk}^2 - M_{ij} M_{ij})$ are the third and second tensor invariants of $M_{ij}$~\cite{MaffettoneMinale98}. The degree of potential droplet deformability is given by the capillary number, which is defined as the ratio of viscous and interfacial forces of the droplet

\begin{equation}
\label{eq:capillary}
\mbox{Ca}(t)=\frac{\eta_{\text{s}} R G(t)}{\sigma}
\end{equation}

where $G(t) \equiv \lVert G_{ij}(t) \rVert_{\infty}$ is the time-dependent shear rate of the solvent flow, $R$ the radius of the initially undeformed droplet, $\sigma$ the surface tension between the droplet and the solvent and $\eta_{\text{s}}$ the dynamic viscosity of the solvent flow. Unlike previous analytical approaches to model droplet deformation~\cite{Taylor1932,Taylor34} the MM model is not based on a perturbative expansion in the capillary number $\mbox{Ca}$. Therefore, the large  $\mbox{Ca}$ regime, in which droplet instabilities and breakup occur, may in theory be described by the MM-model. However, it should be noted, that the MM-model may not give accurate large scale deformation or breakup predictions, as the droplet is assumed to be ellipsoidal at all times. The droplet deformation in the MM-model is given by

\begin{equation}
\label{eq:mm_defo}
D \equiv \frac{L - W}{L + W}
\end{equation}

where $L$ and $W$ are the major and minor ellipsoidal axes respectively. Since we want to compare the LBM simulations with the MM-model, we need to make sure that we remain in the linear flow regime and check that our deformed LBM droplet is actually ellipsoidal at all times. The MM-model includes two parameters $f_1$ and $f_2$, which are dependent on the viscous ratio $\chi = \mu_{\tiny d} / \mu_{\tiny s}$, with $\mu_{\tiny d}$ and $\mu_{\tiny s}$ denoting the dynamic viscosities of the dispersed and solvent phase respectively, and the capillary number $\mbox{Ca}$~\cite{MaffettoneMinale98}:

\begin{align}
\label{eq:f_unbounded}
f_1(\chi) & = \frac{40 \, (\chi + 1)}{(3 + 2 \chi) (16 + 19 \chi)} \notag \\
f_2(\chi, \mbox{Ca}) & = \frac{5}{3 + 2 \chi} + \frac{3 \, \mbox{Ca}^2}{2 + 6 \mbox{Ca}^2}
\end{align}

In addition to the phenomenological MM-model we can use fully resolved Lattice Boltzmann simulations~\cite{Benzi92,Succi01} to study sub-Kolmogorov droplet dynamics and breakup in homogeneous and isotropic turbulent flows. The Lattice Boltzmann Method (LBM) has been extensively used in the field of microfluidics, including extensions to accommodate non-ideal effects~\cite{Sbragaglia12}, coupling with polymer micro mechanics~\cite{Onishi1} and thermal fluctuations \cite{Varnik11,Xue2018}. LBM has also been widely used for the modelling of droplet breakup behaviour~\cite{Komrakova13,Liu,Farokhirad,Onishi2,Xi99,Yoshino08,GuptaSbragaglia2014,GuptaSbragagliaScagliarini2015,Chiappini2018,Chiappini2019}. 
In order to model multicomponent systems with the Lattice Boltzmann Model (LBM) we need to account for interfacial forces between different fluid components. This can be achieved with the Shan-Chen Multi-Component model (SCMC) \cite{Shan93,Shan94}, a diffuse interface model in the framework of the LBM. The hydrodynamical quantities, mass and momentum densities, can then be described as:

\begin{align}
\label{eq:multi_momentum}
\rho(\vec{x}, t) & = \sum_{\sigma} \sum_i g_i^{\sigma}(\vec{x}, t) \notag \\
\rho(\vec{x}, t) \vec{u}(\vec{x}, t) & = \sum_{\sigma} \sum_i g_i^{\sigma}(\vec{x}, t) \vec{c}_i
\end{align}

where $g_i^{\sigma}(\vec{x},t)$ denotes the populations in the LBM model for the fluid component $\sigma$ and $\vec{c}_i$ are the lattice velocities. For example, for a two component system with species $A$ and $B$ the index $\sigma$ can take the values $\sigma = A$ and $\sigma = B$. The interaction at the fluid-fluid interface \cite{Sbragaglia2013,Sega2013} is given by:

\begin{equation}
\label{eq:scmc_interface}
\vec{F}^{\sigma}(\vec{x},t) = - \rho_{\sigma}(\vec{x},t) \sum_{\sigma' \neq \sigma} \sum_{i=1}^{N} \mathcal{G}_{\sigma, \sigma'} w_i \rho_{\sigma'}(\vec{x,t} + \vec{c}_i) \vec{c}_i
\end{equation}

where $\rho_{\sigma}(\vec{x},t)$ is the density field of the fluid component denoted by $\sigma$. $\mathcal{G}_{\sigma,\sigma'}$ is a coupling constant for the two phases $\sigma$ and $\sigma'$ at position $\vec{x}$ and $w_i$ are the lattice isotropy weights. We use the same exact flow boundary conditions as outlined in~\cite{Milan2018}. In order to use arbitrary boundary values of the density $\rho(\vec{x}, t)$ and velocity $\vec{u}(\vec{x},t)$ fields of the solvent fluid we use ghost populations (or halos), which store the equilibrium distribution functions $g_i^{\text{eq}}$ of the boundary density and velocity fields. The equilibrium distribution functions $g_i^{\text{eq}}$ at the domain boundary are given by:

\begin{equation}
\label{eq:equilibrium}
g_i^{\text{eq}} (\vec{x},t) = \rho_b(\vec{x},t) w_i \left ( 1 + 3 \, \vec{c}_i \cdot \vec{u} + \frac{9}{2} (\vec{c}_i \cdot \vec{u})^2 - \frac{3}{2} \vec{u}^2 \right )
\end{equation}
 
with $w_i$ being the lattice weights for the set of lattice vectors $\vec{c}_i$, and $\rho_b(\vec{x},t)$ the density field at the domain boundary. Thus the ghost distributions update the boundary nodes during the LBM streaming step and effectively simulate an exact flow boundary given by the chosen density $\rho_b(\vec{x}, t)$ and velocity $\vec{u}(\vec{x},t)$ fields of the outer fluid~\cite{Milan2018}. The streaming and collision steps are given by the Lattice Boltzmann equation:

\begin{equation}
\label{eq:lbe}
g_i(\vec{x} + \vec{c}_i \Delta t, t + \Delta t) - g_i(\vec{x}, t) = \Omega(\{g_i(\vec{x},t)\})
\end{equation}

where $\Omega(\{g_i(\vec{x},t)\})$ is the collision operator depending on the whole (local) set of lattice populations and $\Delta t$ is the simulation time step. For MRT (multi-relaxation time scale) the collision operator is linear and contains several relaxation times linked to its relaxation modes (depending on the lattice stencil) \cite{DHumieres02}. One relaxation time $\tau$ is directly linked to the kinematic viscosity $\nu$ in the system

\begin{equation}
\label{eq:viscosity}
\nu = \frac{1}{3} \left ( \tau - \frac{1}{2} \right ) 
\end{equation}

which is one of the primary links between the LBM scheme and hydrodynamics \cite{Benzi92,Succi01}. Since the boundary scheme is not strictly mass conserving, we accept small mass fluctuations of both droplet and solvent, but keep the density ratio $\rho_{\tiny \mbox{d}} / \rho_{\tiny \mbox{s}}$ constant, which is achieved by reinjecting mass into the droplet~\cite{Biferale11}.

\section{Simulation setup}
\label{sec:set_up}

\begin{figure}[!htbp]
\centering
\includegraphics[scale=0.4, trim={0mm, 0mm, 0mm, 15mm}, clip]{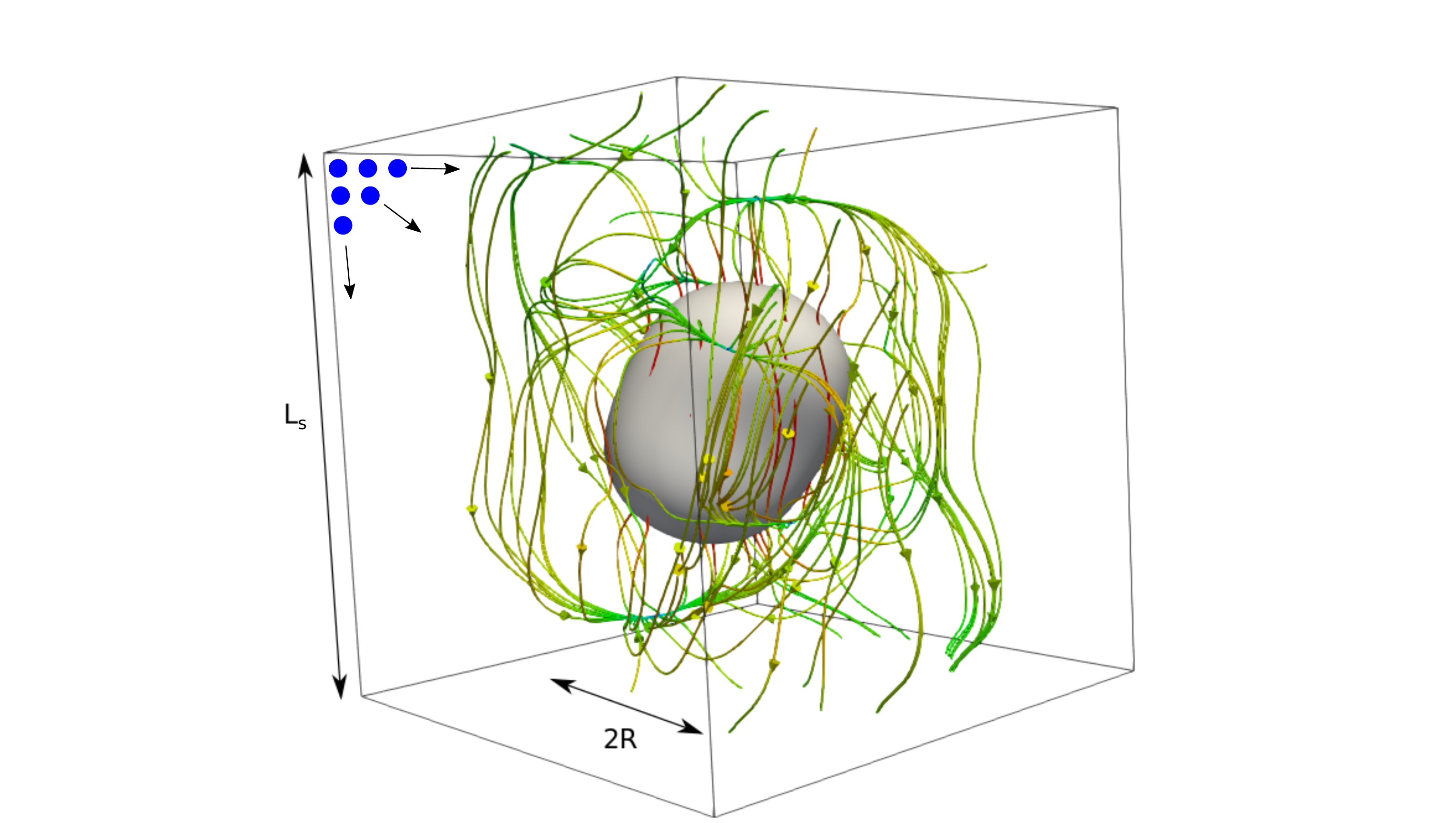}
\caption{LBM simulation setup: the droplet of scale $R$ in the simulation domain with dimensions $L_s \times L_s \times L_s$. The turbulent flow is initialised at the boundary nodes and visualised via streamlines coloured according to the velocity magnitude $\lVert \vec{v} \rVert$. The blue dots indicate the ghost nodes, which are crucial in coupling the pseudo-spectral turbulent flow data to the LBM simulation.}
\label{fig:sketch_droplet}
\end{figure}

\begin{figure}[!htbp]
\centering
\includegraphics[scale=0.75]{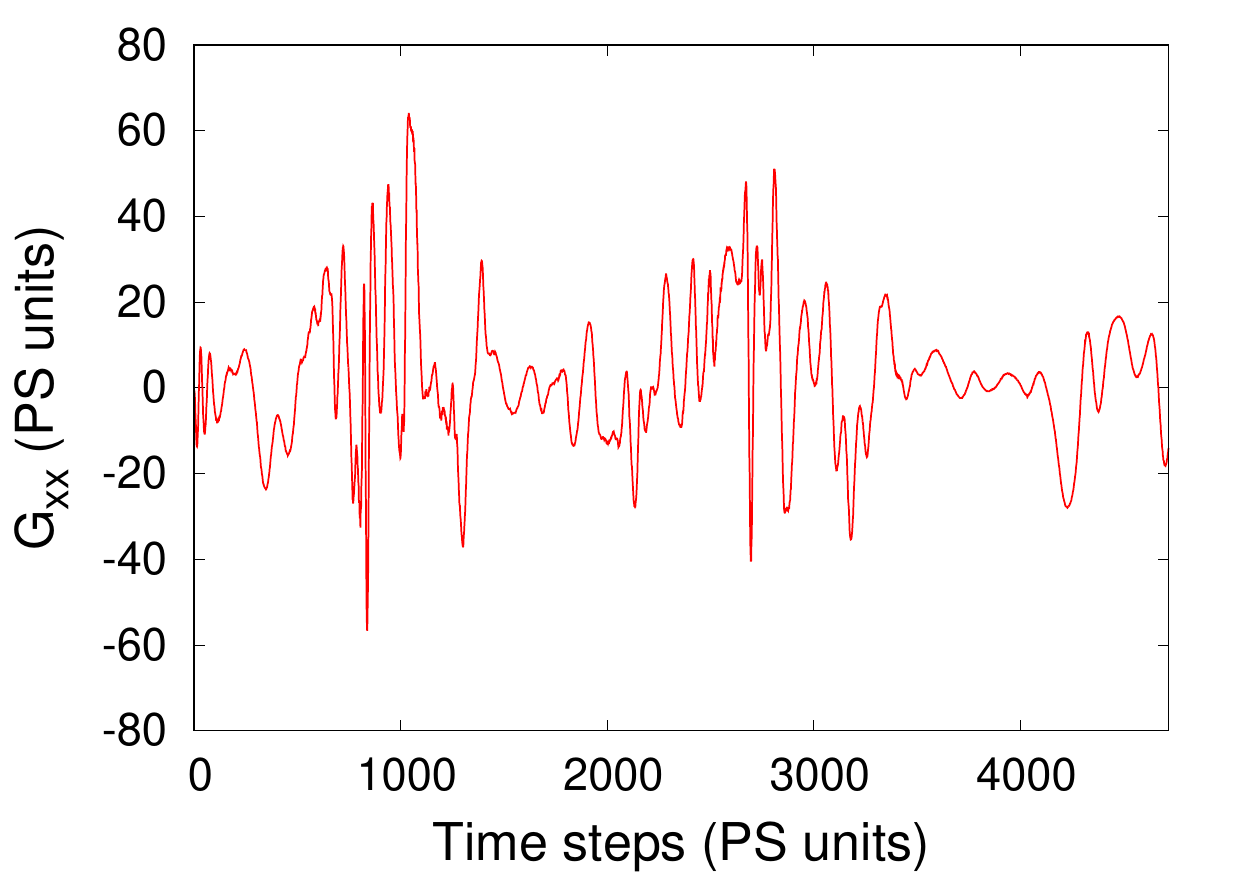}
\caption{The DNS turbulent signal for the component $G_{xx}$ of the velocity gradient tensor $G_{ij}$. The data represents the  $G_{xx}$ velocity gradient component of a tracer trajectory in a homogeneous isotropic turbulent flow obtained via PS simulations.}
\label{fig:dns_grad}
\end{figure}

We have introduced the ghost boundary flow method in section~\ref{sec:set_up} as an exact flow boundary method: the boundary scheme enables us to enforce a density field $\rho_b(\vec{x},t)$ and $\vec{u}(\vec{x},t)$ at the boundaries at any LBM simulation time step $t$. A concrete setup for the sub-Kolmogorov droplet in homogeneous isotropic turbulence is given in figure~\ref{fig:sketch_droplet}. A droplet with radius $R$ in the undeformed stage is at the centre of a simulation box of length $L_s$. The blue dots at the faces of the simulation box denote the ghost nodes, which according to equation~(\ref{eq:equilibrium}) contain the macroscopic boundary values, the density $\rho_b(\vec{x},t)$ and the velocity $\vec{u}(\vec{x},t)$. As in previous chapters we deal with a density ratio of $1$ and thus choose $\rho_b(\vec{x},t)$ accordingly, whereby $\rho_b = \text{constant}$. However, the velocity field $\vec{u}(\vec{x},t)$ is now updated via pseudo-spectral (PS) data of passive trajectories in turbulent flows~\footnote{\url{https://data.4tu.nl/repository/uuid:a64319d5-1735-4bf1-944b-8e9187e4b9d6}}, see figure~\ref{fig:dns_grad}. All the turbulent trajectories have a Taylor Reynolds number of $\mbox{Re}_{\lambda} = 420$. We consider droplets below the Kolmogorov scale, i.e. $R \ll \eta_K$, as we can limit ourselves to a linear velocity profile in this case~\cite{Cristini03}. The Kolmogorov scale is defined as

\begin{equation}
\label{eq:kolmo_scales}
\eta_K \equiv \left ( \frac{\nu_{\text{s}}^3}{\varepsilon} \right )^{\frac{1}{4}}
\end{equation}

where $\nu_{\text{s}}$ is the kinematic viscosity and $\varepsilon$ the energy dissipation of the solvent flow. The typical velocity increments of separation $\vec{l}$ of size $\lvert \vec{l} \rvert = l$ are

\begin{equation}
\label{eq:vel_incremenet}
\delta v_i(\vec{x},\vec{l}) = v_i(\vec{x} + \vec{l}) - v_i(\vec{x})
\end{equation}

which can be approximated by

\begin{equation}
\label{eq:vel_approx}
\delta v_i(\vec{x},\vec{l}) = G_{ij} l_j + \mathcal{O}(l^2)
\end{equation}

for $l \sim R$~\footnote{The authors make use of index notation in equations~(\ref{eq:vel_incremenet}),~(\ref{eq:vel_approx}),~(\ref{eq:linear_turb}) and~(\ref{eq:rescaled_G})}. Since the droplet's centre of mass remains fixed, i.e. we are in the frame of reference of the droplet, we obtain the following velocity relation in the case of a single phase flow:

\begin{equation}
\label{eq:linear_turb}
v_i^{\tiny \mbox{(0)}}(\vec{x},t) = G_{ij}(t) \, x_j
\end{equation}

which is a linear velocity profile in the distance $\vec{x}$ from the droplet's centre of mass. With $v_i^{\tiny \mbox{(0)}}(\vec{x},t)$ we denote the turbulent velocity profile imposed by the boundary, whereas $v_i(\vec{x},t)$ is the LBM obtained velocity profile given by equation~(\ref{eq:multi_momentum}). It is interesting to compare the numerical algorithm for an external homogeneous and isotropic turbulent flow in fully resolved LBM simulations to the study in Mukherjee et al.~\cite{Mukherjee2019}. Instead of an exact boundary flow scheme Mukherjee et al. use a turbulent forcing scheme~\cite{Mukherjee2019,Biferale11} to model the dynamics of droplet emulsions in homogeneous isotropic turbulence. In particular this method has the advantage that the boundary conditions can be chosen as being periodic, which does not require the LBM simulations to converge to a given solution as is the case for the exact boundary method we use. However, the droplets of the turbulent emulsions are larger than the Kolmogorov scale and thus the turbulent forcing scheme in~\cite{Mukherjee2019} is not useful for our study of sub-Kolmogorov droplet dynamics. Following the setup of the exact boundary flow algorithm, we test whether our LBM scheme can relax to the imposed velocity field given in equation~(\ref{eq:linear_turb}).

\section{Validation results}
\label{sec:validation}

Now that we have established the approximations we make to the velocity profile on the sub-Kolmogorov scale, see equation~(\ref{eq:linear_turb}), we would like to create a pseudo-spectral LBM hybrid method, which takes the turbulent velocity gradients $G_{ij}(t)$ as an input. In order for the LBM model to yield accurate predictions, we need to simulate a fully developed turbulent flow at the domain boundaries, which then evolves into the centre of the domain. This is achieved by using both the ghost node boundary flow scheme described in section~\ref{sec:method}, which stores suitable equilibrium distribution functions on ghost nodes, and the pseudo-spectral values for $G_{ij}(t)$. The simulation setup is shown in figure~\ref{fig:sketch_droplet} with the length of the simulation box $L_s$ and the diameter of the undeformed droplet $2 R$ highlighted. 

\begin{figure}[!htbp]
\centering
\includegraphics[scale=0.75]{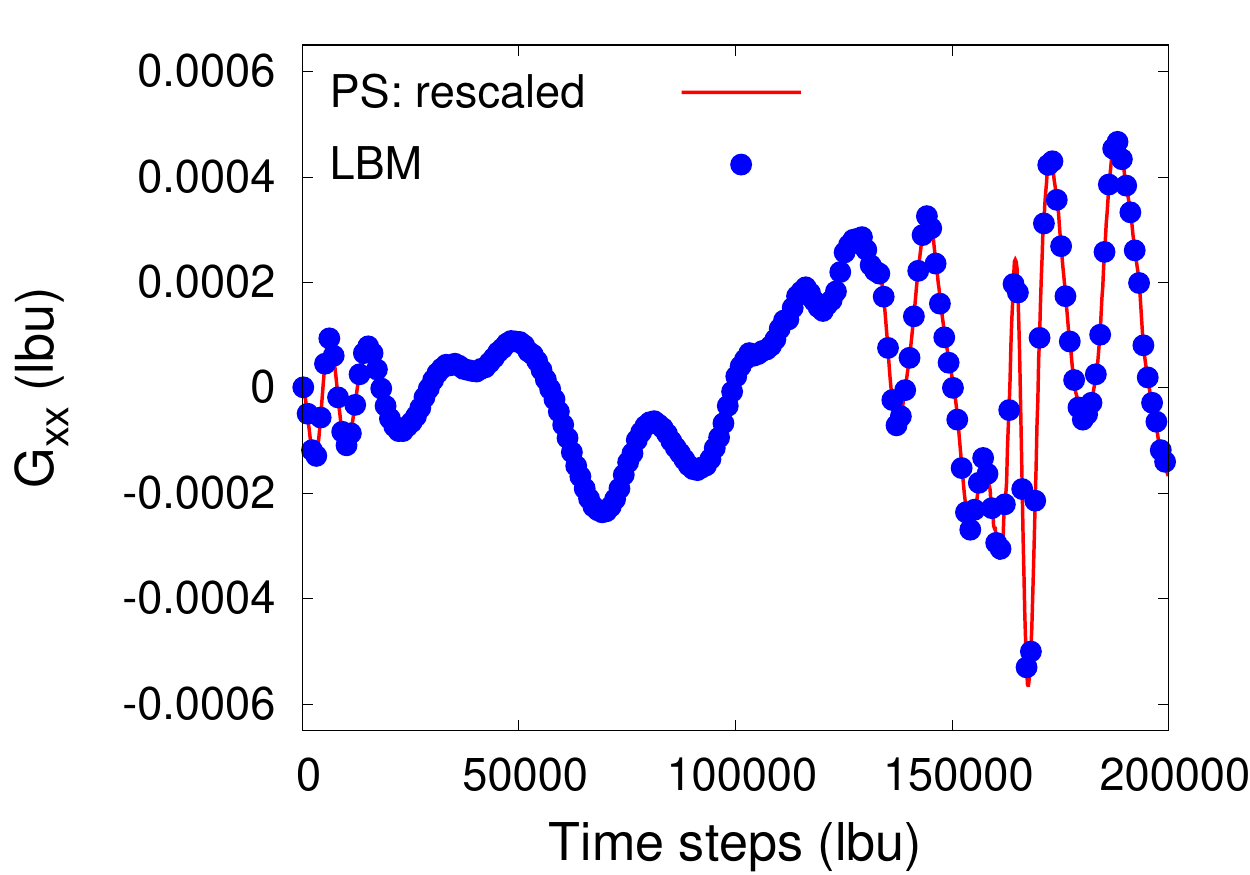}
\caption{Comparison between the velocity gradient of the pseudo-spectral code and the LBM code after rescaling. Since, both curves, the DNS turbulent input data and the LBM reproduced $G_{xx}$, overlap, LBM is able to qualitatively reproduce the turbulent velocity gradient shown in figure~\ref{fig:dns_grad} (only a part of the signal is displayed). The point symbols in the figure are added for a clearer comparison between PS and LBM simulations and do not represent the number of time steps, which is substantially larger.}
\label{fig:turb_grad}
\end{figure}

with the time evolution of the flow completely determined via the turbulent velocity gradient tensor $G_{ij}(t)$. Now we test the convergence of the LBM algorithm for the single phase, where we need to recover the linear profile of the turbulent flow field, see equation~(\ref{eq:linear_turb}), to a given accuracy. Thus, recovering the velocity field in equation~(\ref{eq:linear_turb}) effectively means recovering the turbulent velocity gradient $G_{ij}(t)$ of the pseudo-spectral DNS simulations, as the velocity profile is linear. We choose an LBM sampling time $\Delta t$ measured in lbu, which allows the LBM algorithm to relax to the velocity value required by equation~(\ref{eq:linear_turb}). In figure~\ref{fig:turb_grad}, we see that our LBM scheme reproduces indeed the turbulent signal $G_{ij}(t)$ we provided it with, since the original rescaled DNS signal for the $G_{xx}(t)$ component converges, with the value for $G_{xx}(t)$ obtained with LBM. The rescaling for $G_{ij}(t)$ is 

\begin{equation}
\label{eq:rescaled_G}
G_{ij}(t) \mapsto \frac{u_{\text{\tiny LBM}}}{L_s \lVert G(t) \rVert_{\infty}} G_{ij}(t)
\end{equation}

where $\lVert \ldots \rVert_{\infty}$ is the maximum norm and $u_{\text{\tiny LBM}}$ is a typical LBM velocity scale for which the Mach number $\mbox{Ma} \leq 0.2$. In order to choose the parameter $\Delta t$ we need to check the global convergence of the LBM velocity field $\vec{v}(\vec{x},t)$ with the one imposed at the boundary, see equation~(\ref{eq:linear_turb}). The corresponding $L_2$-error is provided in figure~\ref{fig:turbulence_convergence} and is defined via:

\begin{figure}[!htbp]
\centering
\includegraphics[scale=0.75]{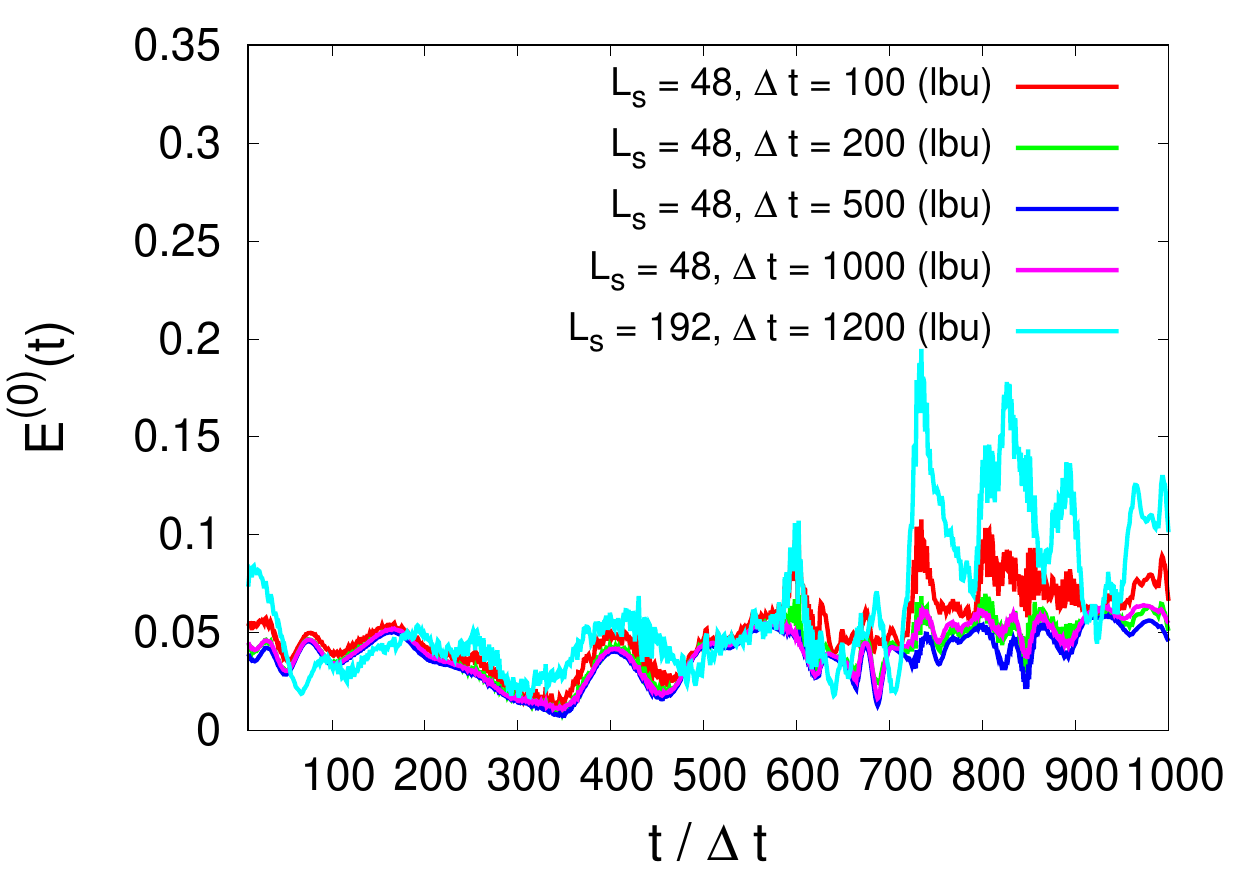}
\caption{Convergence of the time dependent LBM boundary scheme for a single phase homogeneous and isotropic turbulent flow (domain size: $L_s \times L_s \times L_s$). The time averaged global error can be estimated to be about $ \langle E^{\tiny \mbox{(0)}} \rangle_t \sim 0.05$ for a chosen range of $\Delta t$ in terms of LBM time steps $t$ (in lbu) for $L_s = 48$. For larger $L_s$, e.g. $L_s = 192$, $E^{\tiny \mbox{(0)}}(t)$ can take higher values of up to $0.2$. Therefore, the tracer trajectory in use for large $L_s \sim 100$ need to be analysed individually, as the global error convergence may contain large fluctuations.}
\label{fig:turbulence_convergence}
\end{figure}

\begin{equation}
\label{eq:error_turbulent}
E^{\tiny \mbox{(0)}}(t) = \frac{1}{u_{\text{rms}}^{\tiny \mbox{(0)}}(t)} \left [ \frac{1}{L_s^3} \, \int_{0}^{L_s} dx \int_{0}^{L_s} dy \int_{0}^{L_s}  dz \, (\vec{v}(x, y, z, t) - \vec{v}^{\tiny \mbox{(0)}}(x, y, z, t))^2 \right ]^{\frac{1}{2}}
\end{equation}

where $\vec{v}(x, y, z, t)$ and $\vec{v}^{\tiny \mbox{(0)}}(x, y, z, t))$ are the velocity fields of the LBM simulation and the imposed turbulent linear velocity profile respectively. $u_{\text{rms}}^{\tiny \mbox{(0)}}(t)$ is the rms value of $\vec{v}^{\tiny \mbox{(0)}}(x, y, z, t)$. We would like to choose a minimal $\Delta t$, as to optimise run time, we choose $\Delta t = 200 \, (\text{lbu})$, since this value yields a time averaged global error  of $\langle E^{\tiny \mbox{(0)}} \rangle_t < 0.1$. This is a reasonable error threshold, since LBM simulations for droplet dynamics in oscillatory shear flows yield a similar maximal error~\cite{Milan2018}. However, we should remark, that the global error $E^{\tiny \mbox{(0)}}(t)$ in figure~\ref{fig:turbulence_convergence} for a larger system size of $L_s = 192$ has peak values of about $E^{\tiny \mbox{(0)}}(t) \sim 0.2$. This is a significant error to the theoretical linear profile given in equation (~\ref{eq:linear_turb}). Nevertheless, we can still employ the exact flow boundary scheme for the dynamics of sub-Kolmogorov droplets, if we check the bahviour of the error function $E^{(0)}(t)$ for a specific trajectory in use in the single phase case, before setting up the droplet simulations. In case the error function $E^{(0)}(t)$ of the single phase velocity $\vec{v}(x, y, z, t)$ is below 0.1 for every $t < t_{\text{max}}$, where $t_{\text{max}}$ indicates the time where the maximum of the trajectory is reached, the external turbulent flow will be (almost) linear and we obtain an accurate simulation of a turbulent flow corresponding to the trajectory in use.

\begin{table*}[!htbp]
\centering
\begin{tabular}{ @{} c c c c c @{} }
\toprule[1pt]
$L_s$ (lbu) & $\tau$ (lbu) & $u_{\text{\tiny LBM}}$ (lbu) & $\Delta t$ (lbu)  \\
\midrule[0.5pt]
\addlinespace[0.5mm]
48 & 1.0 & 0.1 & 100  \\
48 & 1.0 & 0.1 & 200  \\
48 & 1.0 & 0.1 & 500  \\
48 & 1.0 & 0.1 & 1000  \\
192 & 1.0 & 0.1 & 1200  \\
\bottomrule[1pt]
\end{tabular}
\caption{The simulation parameters for the error function of the global velocity profile $E^{(0)}(t)$ in figure~\ref{fig:turbulence_convergence}.}
\label{tab:validation}
\end{table*}

\section{Simulation results}
\label{sec:results}

Since we have established the validity of the hybrid PS-LBM algorithm in section~\ref{sec:validation}, we can investigate droplet dynamics and breakup for a single droplet on the sub-Kolmogorov scale. Analogously to equation~(\ref{eq:mm_defo}) we can define a measure for the deformation $D_\text{LBM}$ for our LBM simulations:

\begin{equation}
\label{eq:lbm_defo}
D_{\text{LBM}}(t) = \sqrt{1 - \frac{\Omega_0(t)}{\Omega(t)}}
\end{equation}

\begin{figure}[!htbp]
\centering
\includegraphics[scale=1.0]{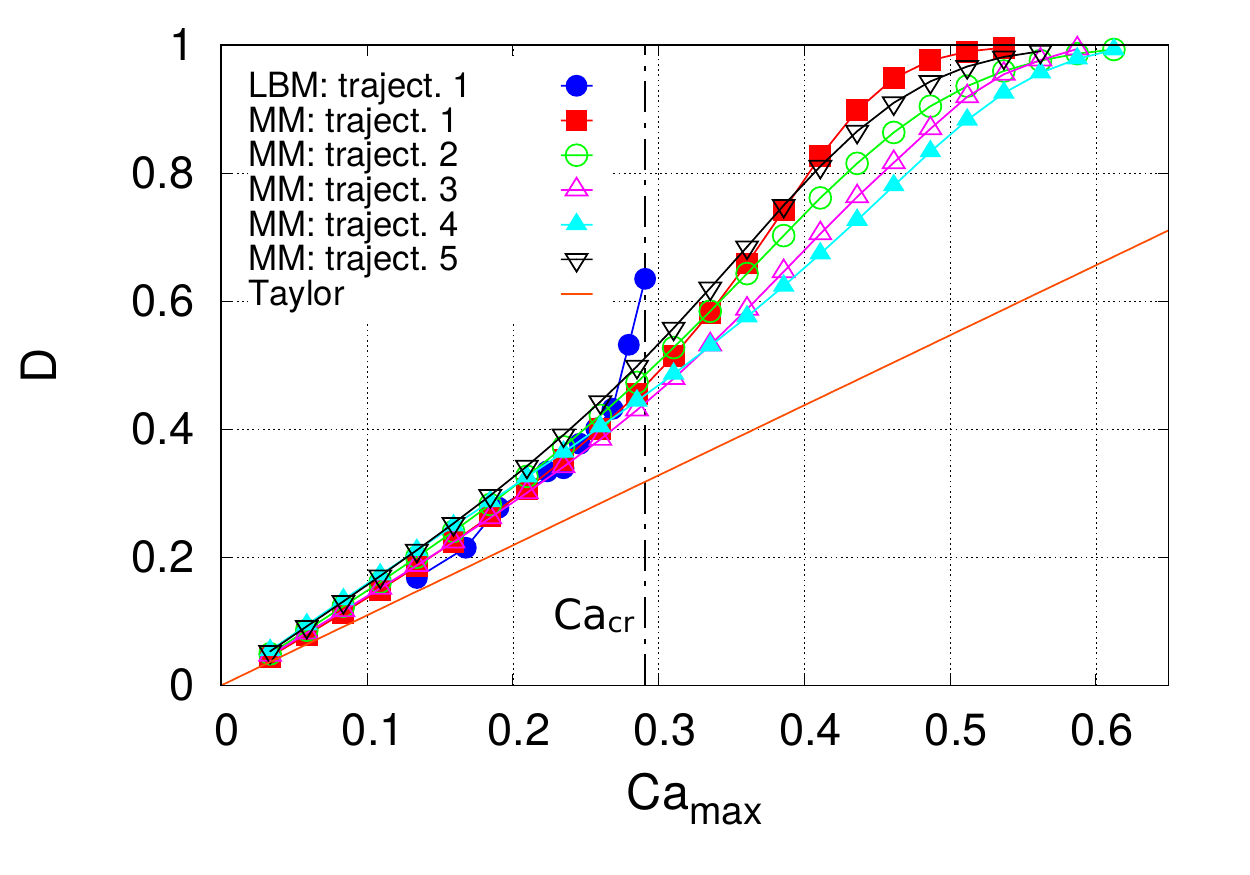}
\caption{MM-model predictions for the ellipsoidal droplet deformation $D$ as a function of $\mbox{Ca}_\text{max}$ for different  turbulent trajectories. The critical capillary number $\mbox{Ca}_\text{cr}$ is estimated via a cut off for $D$ in the MM-model. The Taylor deformation law~\cite{Taylor1932,Taylor34} is shown for comparison in the low $\mbox{Ca}_\text{max}$ limit. We can see that the LBM results coincide with the one of the MM-model, up until a maximum capillary number $\mbox{Ca}_\text{max} \approx 0.291$, where the droplet is no longer ellipsoidally deformed, which causes the deviations from the MM-model predictions, see figures~\ref{fig:droplet_breakup} and~\ref{fig:droplet_breakup_streamlines}. The critical capillary number obtained via LBM is $\mbox{Ca}_\text{cr} = 0.286 \pm 0.006$, which is lower than the one predicted in the MM-model, which is lies in the range $\mbox{Ca}_\text{cr}^{\text{MM}} = 0.58 \pm 0.05$. Therefore, it appears that the MM-model is not able to predict the correct critical capillary number $\mbox{Ca}_\text{cr}$ for sub-Kolmogorov droplet breakup.}
\label{fig:mm_turb}
\end{figure}

\begin{figure}[!htbp]
\centering
\vcenteredhbox{\includegraphics[scale=0.25]{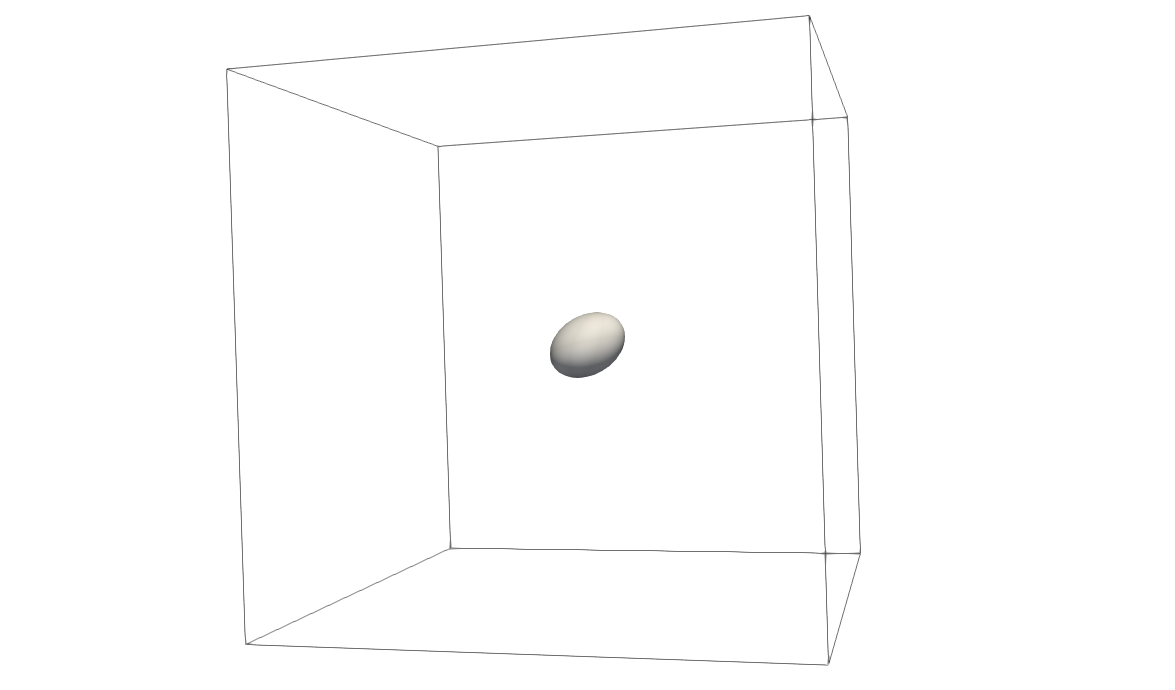}}
\vcenteredhbox{\includegraphics[scale=0.25]{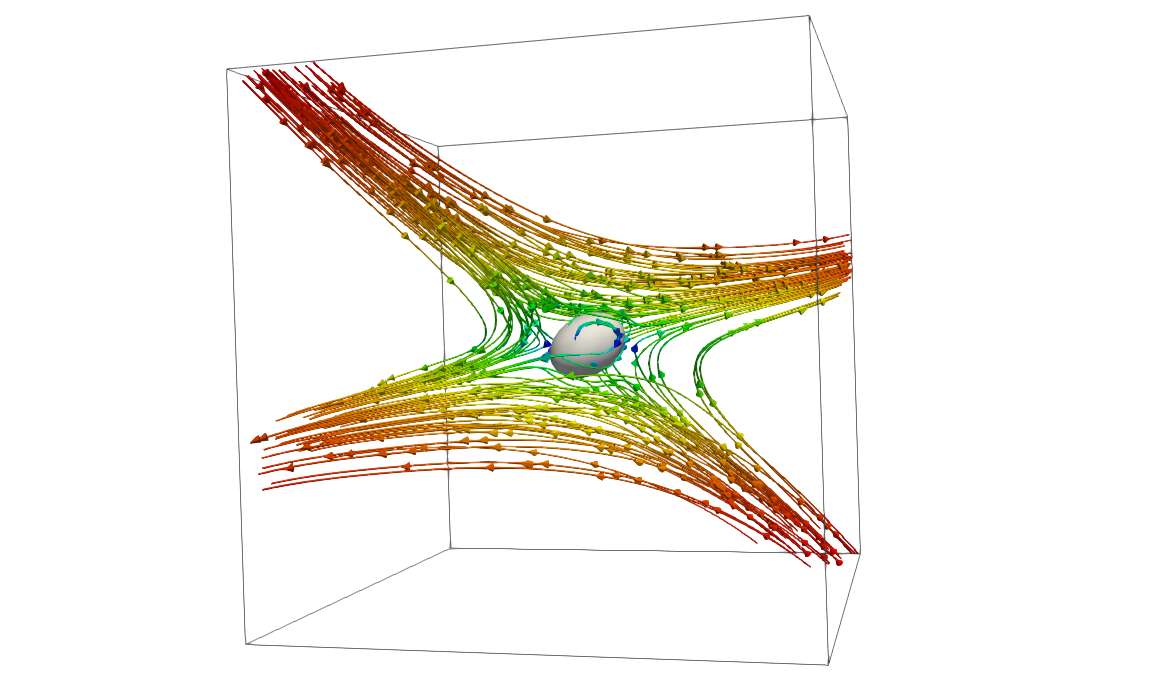}}
\caption{Sub-Kolmogorov droplet for a capillary number $\mbox{Ca}_\text{max} \ll \mbox{Ca}_{\text{cr}}$ with and without the velocity field shown via streamlines. The droplet is ellipsoidally deformed, which is the regime where LBM and MM predictions coincide.}
\label{fig:droplet_flow}
\end{figure}

\begin{figure}[!htbp]
\centering
\vcenteredhbox{\includegraphics[scale=0.25]{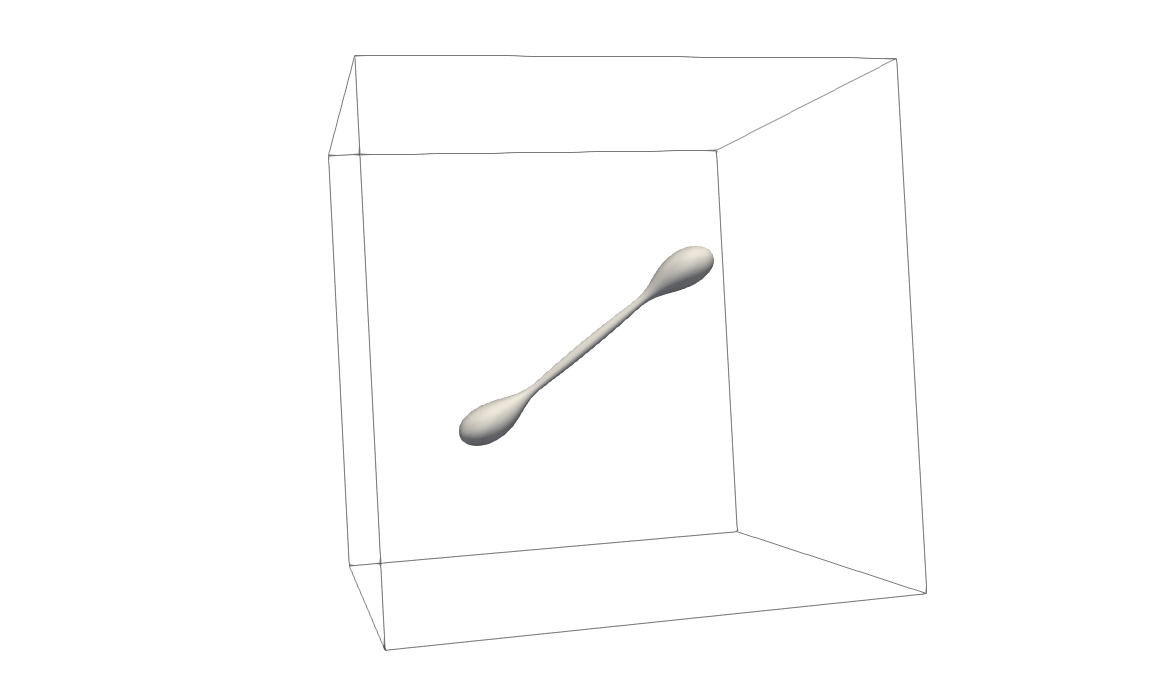}}
\vcenteredhbox{\includegraphics[scale=0.25]{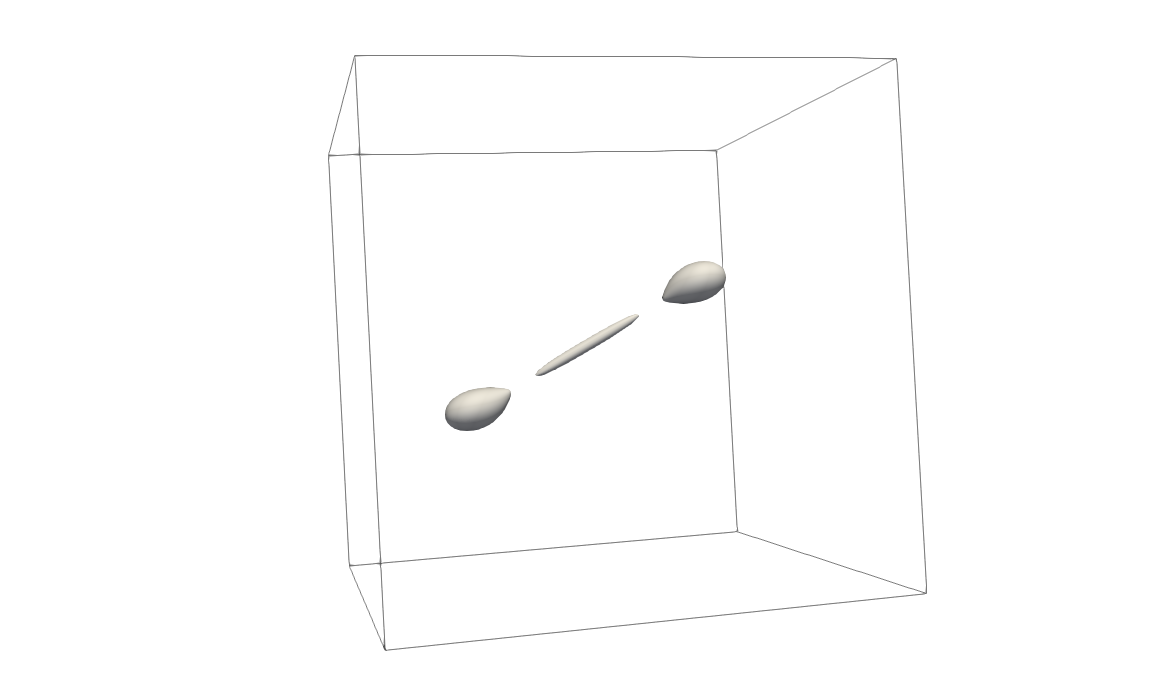}}
\caption{Sub-Kolmogorov droplet for a capillary number $\mbox{Ca}_\text{max} \approx \mbox{Ca}_{\text{cr}}$ before and after breakup. The elongated droplet shape indicates that substantial deviations from the MM-model predictions might be possible.}
\label{fig:droplet_breakup}
\end{figure}

\begin{figure}[!htbp]
\centering
\vcenteredhbox{\includegraphics[scale=0.25]{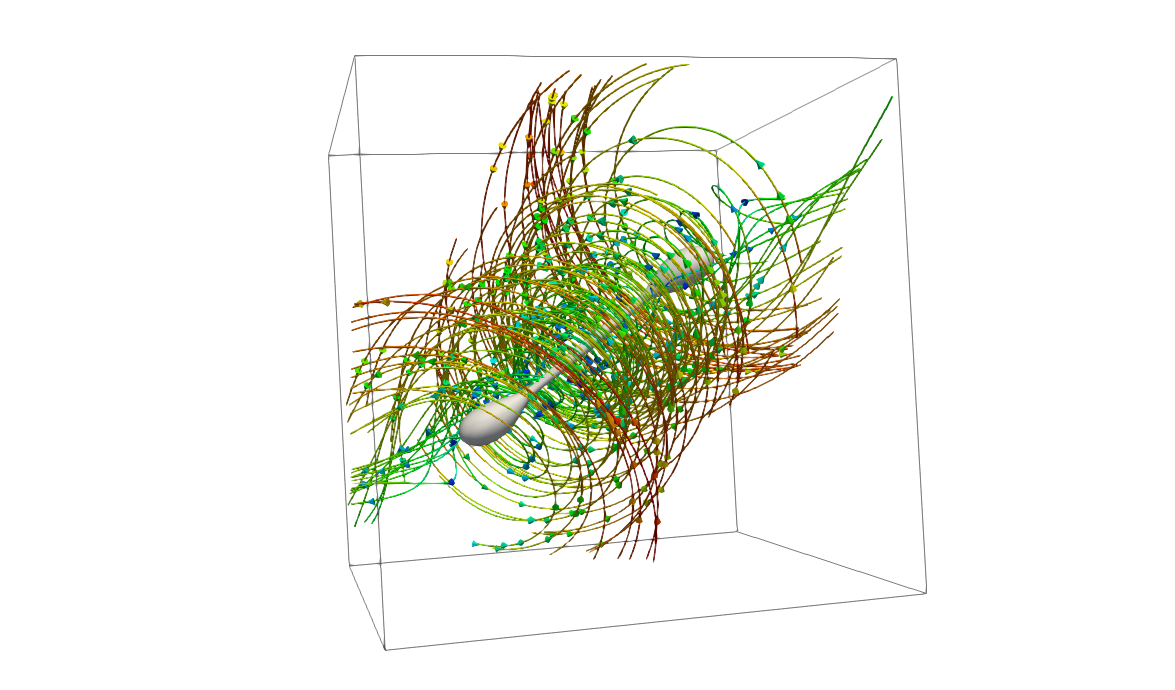}}
\vcenteredhbox{\includegraphics[scale=0.25]{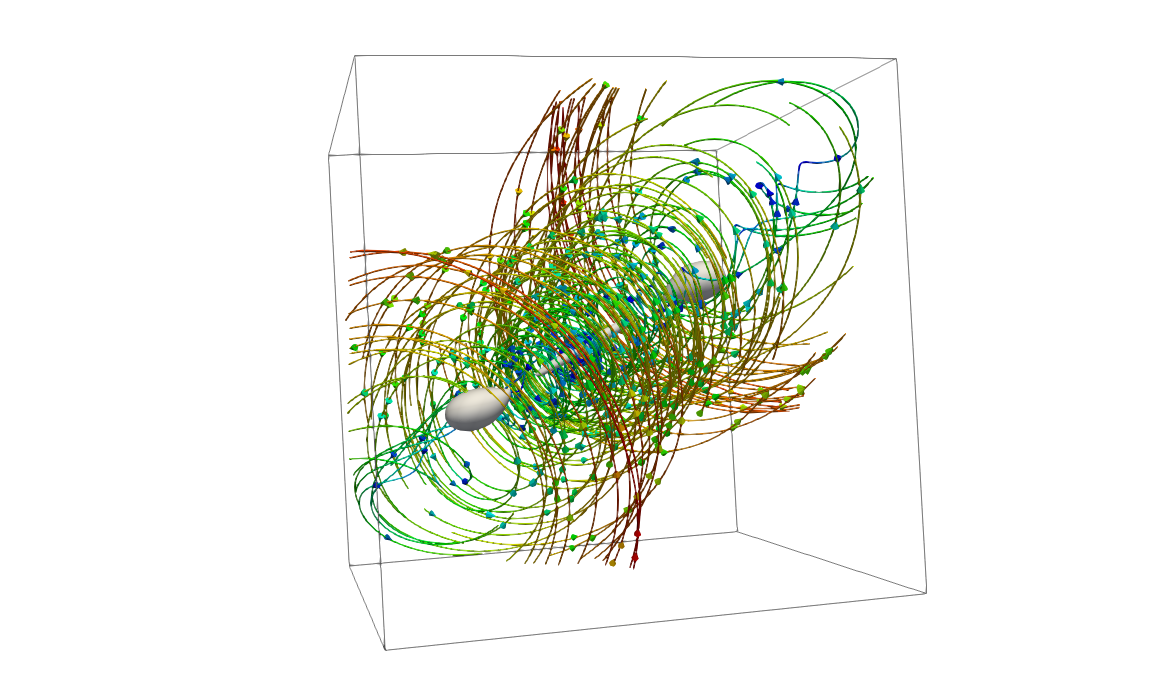}}
\caption{Sub-Kolmogorov droplet with the velocity field indicated via streamlines for a capillary number $\mbox{Ca}_\text{max} \approx \mbox{Ca}_{\text{cr}}$ before and after breakup. We see that the elongated shape of the droplet is due to both a strong strain and rotational part in the turbulent flow field at this time instance.}
\label{fig:droplet_breakup_streamlines}
\end{figure}

\begin{table*}[!htbp]
\centering
\begin{tabular}{ @{} c c c c c @{} }
\toprule[1pt]
$L_s$ (lbu) & $\tau$ (lbu) & $R$ (lbu) & $\Delta t$ (lbu)   \\
\midrule[0.5pt]
\addlinespace[0.5mm]
288 & 3.0 & 20 & 200 \\
\bottomrule[1pt]
\end{tabular}
\caption{LBM simulation parameters for the droplet deformation in figure~\ref{fig:mm_turb}. The relaxation time $\tau$ is the same for both the dispersed and solvent phase.}
\label{tab:results}
\end{table*}

where $\Omega(t)$ is the time-dependent surface area of the deformed droplet and $\Omega_0(t)$ is the surface area of an undeformed spherical droplet of the same volume as the deformed droplet. $D_{\text{LBM}}(t)$ represents a bounded measure for droplet deformation, because $D_{\text{LBM}}(t) = 0$ in case of minimal deformation, for which $\Omega(t) = \Omega_0(t)$, and $D_{\text{LBM}}(t) \to 1$ for $\Omega(t) \to \infty$ in case of a hypothetically arbitrary large surface area $\Omega(t)$. It should be noted that $D_{\text{LBM}}(t)$ scales with the droplet length $L(t)$ for elongated droplets~\cite{Milan2020} and thus $D_{\text{LBM}}(t) \sim D$, where $D$ is the droplet deformation according to major and minor ellipsoidal axes, see equation~(\ref{eq:mm_defo}). We start our simulations by small amplitude deformations, so that the droplet is only ellipsoidally deformed, for which we expect the LBM and MM-model results to match, see figure~\ref{fig:droplet_flow}. The maximum capillary number $\mbox{Ca}_\text{max}$ between different runs is changed via the LBM velocity sacle $u_{\text{\tiny LBM}}$, see equation~(\ref{eq:rescaled_G}). As we can see from figure~\ref{fig:mm_turb}, this is indeed the case, because both predictions for the deformation $D$, LBM and the MM-model, coincide for a maximum capillary number $\mbox{Ca}_\text{max} \ll \mbox{Ca}_{\text{cr}}$. Figure~\ref{fig:mm_turb} shows the deformation curves for two different turbulent tracer trajectories for both the hybrid DNS LBM scheme and the MM-model, labelled trajectory $1$. In addition, we plot further four MM-model deformation curves to be able to qualitatively compare both deformation curves from different tarcer trajectories and the MM-model predictions with the one of the LBM hybrid scheme. The MM-model deformation curves and the LBM curve collapse for the intermediate range of $\mbox{Ca}_\text{max}$ until breakup occurs for the LBM curve at around $\mbox{Ca}_\text{max} \approx 0.291$, see figures~\ref{fig:droplet_breakup} and~\ref{fig:droplet_breakup_streamlines}. Figure~\ref{fig:droplet_breakup} shows the elongated droplet for a maximal capillary number $\mbox{Ca}_\text{max} \approx \mbox{Ca}_{\text{cr}}$, just before and after breakup. Interestingly, the droplet is now elongated and thus resembles droplets close to $\mbox{Ca}_{\text{cr}}$ in a pinch-off breakup process found in studies on both confined droplets in laminar flows~\cite{Vananroye08,Vananroye11b,Janssen10,Milan2020} and computational studies on breakup of sub-Kolmogorov droplets~\cite{Cristini03,Khismatullin03,Komrakova13}, and breakup of droplets in porous media flows~\cite{Patel03}. Figure~\ref{fig:droplet_breakup_streamlines} shows the same droplet breakup as figure~\ref{fig:droplet_breakup_streamlines} with the velocity field shown via streamlines. We can see that the turbulent flow possesses a significant rotational part in the instance of the breakup. Let us consider the deformation diagram of figure~\ref{fig:mm_turb} again: LBM and the MM-model predict two very different ranges for the critical capillary number $\mbox{Ca}_{\text{cr}}$. Firstly, this is due to the fact, that breakup in the MM-model is determined via a cut off in the deformation $D$, as the model does not account for a breakup mechanism per se. Even if the cut off is changed slightly, the MM-model still gives similar breakup predictions. Secondly, the MM-model can only model ellipsoidally deformed droplets, even in the case of large amplitude flows. Therefore, the MM-model is unable to model the elongated droplet of the DNS-LBM hybrid simulations, shown in figure~\ref{fig:droplet_breakup}, and thus cannot give a very accurate prediction for the critical capillary number $\mbox{Ca}_{\text{cr}}$. The MM-model prediction for the critical capillary number is $\mbox{Ca}_\text{cr}^{\text{MM}} = 0.58 \pm 0.05$, where the error has been estimated via the spread of the MM deformation curves in figure~\ref{fig:mm_turb}. LBM predicts a lower range of $\mbox{Ca}_\text{cr} = 0.286 \pm 0.006$, which is more accurate due to the fully resolved LBM SCMC model, even though the error might be larger if more turbulent flow trajectories were considered, as has been the case for the MM-model in figure~\ref{fig:mm_turb}. The LBM error in $\mbox{Ca}_\text{cr}$ is estimated via a step in the maximal critical capillary number $\Delta \mbox{Ca}_\text{max} \approx 0.011$. $\Delta \mbox{Ca}_\text{max}$ represents the change in capillary number between individual data points in figure~\ref{fig:mm_turb}. Interestingly, the LBM result of $\mbox{Ca}_\text{cr} = 0.286 \pm 0.006$ for the critical capillary number are in very good agreement with the results reported in~\cite{Khismatullin03} and~\cite{Komrakova13}, for $\mbox{Re} = 1$ and $\chi = 1$.

\section{Conclusion}
\label{conclusion}
The DNS-LBM hybrid scheme for sub-Kolmogorov droplets in fully developed homogeneous and isotropic turbulence has been used to model droplet deformation and breakup. The LBM results were also compared to MM-model predictions. We have found that LBM predicts a significantly lower critical capillary number $\mbox{Ca}_{\text{cr}}$ than the MM-model, see figure~\ref{fig:mm_turb}. The reason for this is probably two-fold: firstly breakup is chosen via a cut off procedure for the deformation parameter $D$ in the MM-model and secondly the droplet is far from an ellipsoidal shape in the region of maximum capillary numbers $\mbox{Ca}_{\text{max}} \leq \mbox{Ca}_{\text{cr}}$, see figures~\ref{fig:droplet_breakup} and~\ref{fig:droplet_breakup_streamlines}. This suggests, that even though the MM-model and DNS-LBM predictions agree very well in the capillary number region $\mbox{Ca}_{\text{max}} \ll \mbox{Ca}_{\text{cr}}$, accurate predictions of sub-Kolmogorov droplet breakup seem to be only viable with fully resolved simulations, such as the DNS-LBM hybrid algorithm, instead of phenomenological models such as the MM-model~\cite{Biferale2014,Ray2018}. It would be interesting to investigate deformation statistics of Sub-Kolmogorov droplets extensively with the DNS-LBM hybrid algorithm and once again use the MM-model for comparison~\cite{Biferale2014,Ray2018}. In addition, it would be interesting to see the range of critical capillary numbers  $\mbox{Ca}_{\text{cr}}$ predicted by the DNS-LBM hybrid model by extending the LBM results of droplet breakup in figure~\ref{fig:mm_turb} for a large number of turbulent tracer trajectories.

\section*{Acknowledgements}

The authors kindly acknowledge funding from the European Union's Framework Programme for Research and Innovation Horizon 2020 (2014 - 2020) under the Marie 
Sk\l{}odowska-Curie Grant Agreement No.642069 and funding from the European Research Council under the European Community's Seventh Framework Program, ERC Grant Agreement No 339032. The authors would also like to thank Fabio Bonaccorso, Dr Anupam Gupta, Dr Gianluca di Staso and Xiao Xue for their support.

\section*{Author contribution statement}

All of the authors were involved in the preparation of the manuscript and have read and approved the final manuscript version.

\bibliographystyle{elsarticle-num}
\bibliography{bibfile}

\end{document}